\documentclass[aps,preprint,groupedaddress,showpacs]{revtex4-1}
\usepackage{amsmath}
\usepackage{amssymb}
\usepackage{graphicx}
\usepackage{dcolumn}

\usepackage{times}
\usepackage[colorlinks,linkcolor=blue,citecolor=blue,urlcolor=blue]{hyperref}
\usepackage{natbib}

\mathchardef\mhyphen="2D

\begin{document}

\author{M. Sch\"uler}
\author{Y. Pavlyukh}
\author{J. Berakdar}
\altaffiliation{Author to whom any correspondence should be addressed}

%
\address{Institut f\"ur Physik, Martin-Luther Universit\"at Halle Wittenberg,
  Halle, Heinrich-Damerow-Stra\ss e 4, 06120, Germany}
\email{jamal.berakdar@physik.uni-halle.de}
\title{Theory of  spin dynamics of magnetic adatoms traced by time-resolved scanning tunneling
spectroscopy}

\date{\today}

\begin{abstract}


The inelastic scanning tunneling microscopy (STM) has been shown recently (Loth \emph{et
al.} Science \textbf{329}, 1628 (2010)) to be extendable as to access the nanosecond,
spin-resolved dynamics of magnetic adatoms and molecules.  Here we analyze theoretically
this novel tool by considering the time-resolved spin dynamics of a single adsorbed Fe
atom excited by a tunneling current pulse from a spin-polarized STM tip.  The adatom
spin-configuration can be controlled and probed by applying voltage pulses between the
substrate and the spin-polarized STM tip.  We demonstrate how, in a pump-probe manner, the
relaxation dynamics of the sample spin is manifested in the spin-dependent tunneling
current.  Our model calculations are based on the scattering theory in a wave-packet
formulation. The scheme is nonpertubative and hence, is valid for all voltages. The
numerical results for the tunneling probability and the conductance are contrasted with
the prediction of simple analytical models and compared with experiments.

\end{abstract}


\maketitle
\tableofcontents

\section{Introduction}

Spin systems are extensively utilized as essential elements for (quantum) information
storage or computing devices~\cite{spincomputing1, spincomputing2}.  Thereby,
environmental effects play an important role as they generally lead to decoherence and
relaxation.  The time scales for these processes depend on the underlying coupling
mechanisms and exhibit in general a marked temperature and dimensionally dependence.  A
detailed insight and a possible control of this relaxation is a key factor for the
operation of these devices.  Desirably the relaxation time should be larger than the
operation (switching) time.  In this sense, it is of importance to identify spin systems
with suitable relaxation properties and amenable to full fast control of the
spins. Prototypical examples are realized in semiconductor
nanostructures~\cite{magnetsemicond,semiconductorspintronics} or
nanomagnets~\cite{molecularmagnets1,molecularmagnets2}. For a single magnetic atom or
molecules adsorbed on a substrate, the surface spin excitations are usually weakly coupled
to bulk magnons resulting in relaxation times on the order of nanoseconds, which has been
demonstrated to be sufficient for a coherent spin
manipulation~\cite{coherentmanipulation}.

The experimental analysis of surface-deposited structures can be performed conveniently by
means of the scanning tunneling microscopy, which has an advantage of an atomic spatial
resolution. Using a spin-polarized tip, the spin of a single magnetic adatom can be probed
and manipulated~\cite{experiment}, since the tunneling probability and the current depend
on the relative alignment of the surface and the tip magnetic moments. Very recently the same
group demonstrated the potential of the technique for atomic-scale information storage and
retrieval~\cite{experiment2}.  The time resolved STM experiments~\cite{timeresolvedstm}
renders possible the observation of the quantum dynamics. The transient precessional
dynamics of the excited spin states is still too fast to be measured.  The relaxation
process can, however, be mapped onto the current-dependence in a pump-probe
experiment. Thus one has a possibility to directly measure the relaxation time of such
systems which opens the way for testing different configurations with maximal coherence time.

To our knowledge, the first experiment to measure the relaxation times via STM
 has been carried out by S. Loth
\emph{et.~al.} (Ref.~\onlinecite{experiment}) for a  magnetic system
with a particularly long spin relaxation time of above $200$ ns.
The basic experimental setup
(figure~\ref{setup}) consists of a Fe-Cu dimer  placed on the Cu (100) surface covered
with a Cu$_2$N overlayer and then probed by a spin-polarized tip. A magnetic field of
$B=7$ T is applied, almost aligning the spin of the Fe atom parallel to the spin axis of
the tip. Due to the magnetic coupling between the spin of the tunneling electrons and the
spin of the Fe atom, the adsorbate is driven into an excited state with a different
projection of the spin moment on the spin axis of the tip. This influences the tunneling
current and allows to trace the spin dynamics.

We will consider theoretically the experimental situation (section \ref{secmodel}) using
scattering theory (section~\ref{secscattering}) with the aim to calculate the
spin-dependent transport properties. The method is complementary to other formalisms. A
widely spread approach is to describe the tunneling by the master equation with transition
rates calculated in a standard ``golden rule'' manner~\cite{Recher2000,Delgado2011}. This
method is, in essence, a perturbative formalism and hence requires a separate
justification for each of the considered experimental configurations. The nonequilibrium
Green's function approach has a potential to systematically take into account many-body
effects, however, the transition rates are often introduced as adjustable
parameters~\cite{Penteado2011}. In the present method we focus on the description of the
tunneling starting from a model Hamiltonian including the magnetic anisotropy, the
exchange coupling and the coupling to the external magnetic field.  Our theory proves that
it is indeed possible to directly measure the spin-relaxation times provided they are
longer than the precession time. Furthermore, our main findings for the relative number of
transmitted electrons with respect to the pump-probe delay (presented in section
\ref{secpumpprobe}) allow for a rigorous determination of the magnetic coupling
parameters.

\section{Model Hamiltonian \label{secmodel}}

\begin{figure}
\includegraphics[width=0.45\columnwidth]{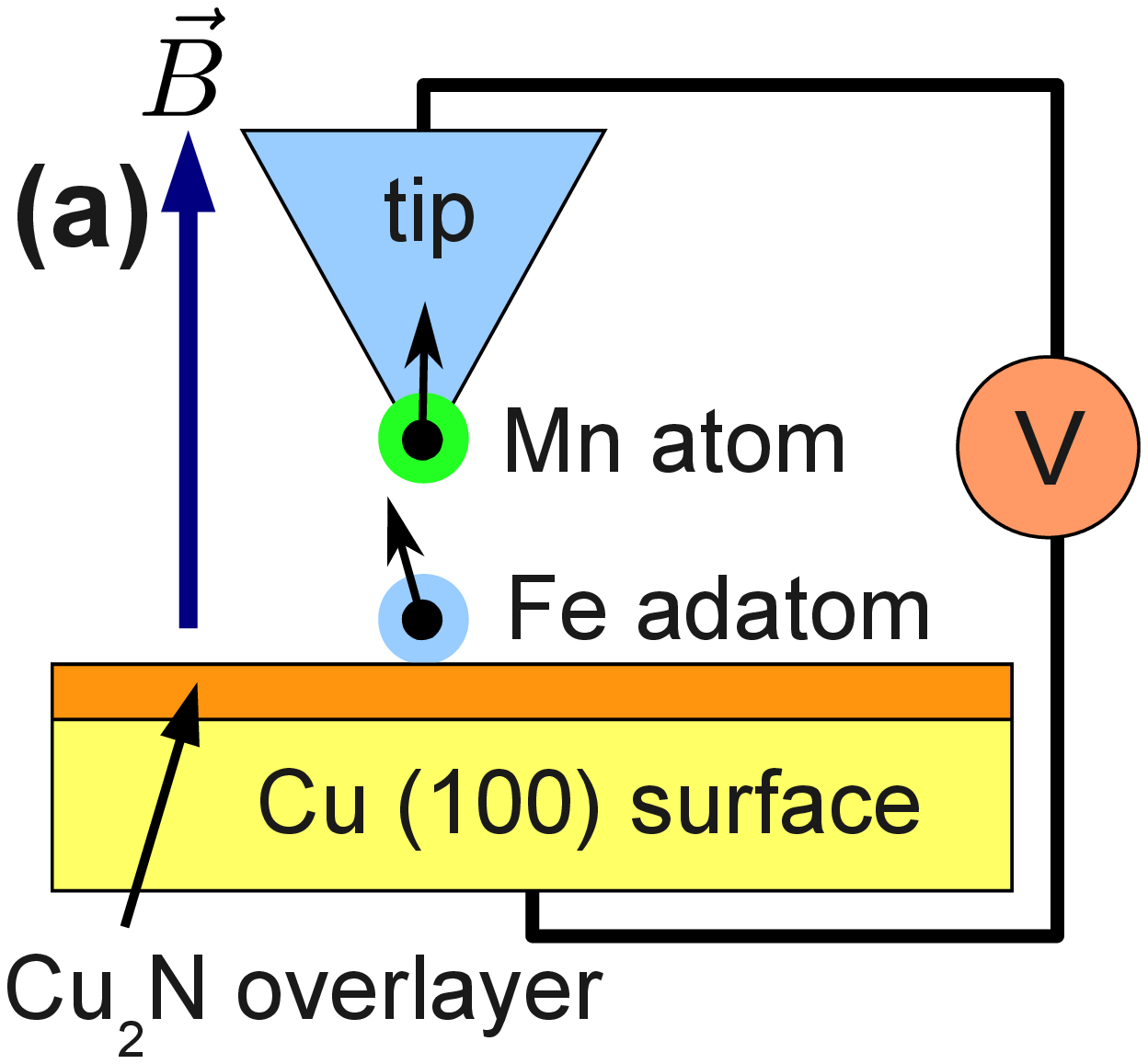}
\includegraphics[width=0.5\columnwidth]{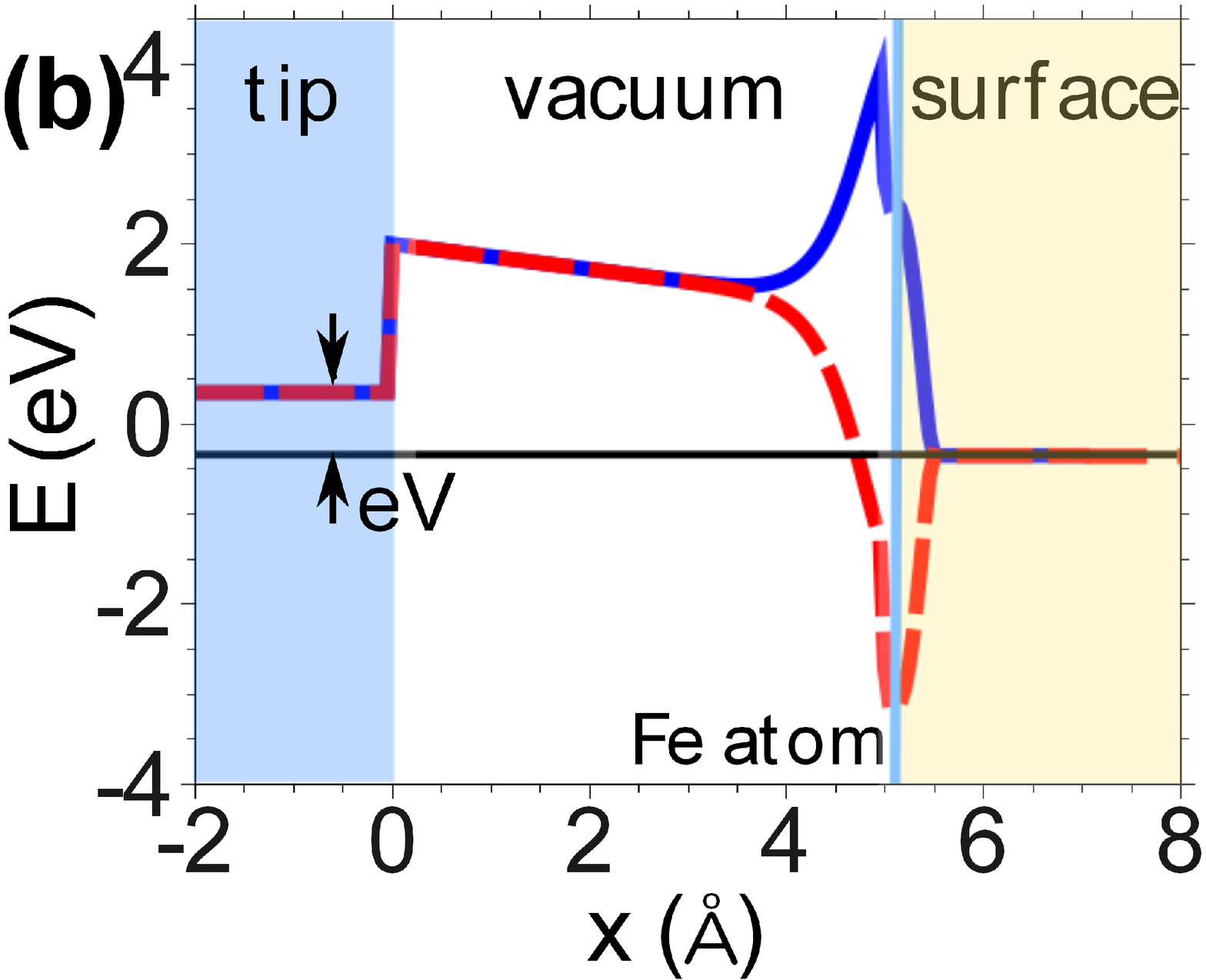}
\caption{Color online (a) A spin-polarized STM tip is placed above a single Fe adatom
adsorbed on the copper (100) surface with an Cu$_2$N overlayer to excite and probe the
spin dynamics of the adatom. The tip consists of  one Mn atom attached at the end of a tungsten tip.
The  Mn atom  acts as a spin filter for the electrons propagating from the tip to
the sample. A strong magnetic field of $B = 7$ T adjusts the spin projection of the Fe atom nearly
parallel to the spin polarization direction of the tip. (b) A theoretical model of (a) by
a tunneling barrier and a spin-dependent interaction potential around the Fe atom on the
surface. The solid line corresponds to the surface-deposited   ground state spin.  The
coupling between surface-deposited  spin and tip-spin  states is antiferromagnetic. Hence,  the coupling energy
decreases when exciting the Fe spin states. The effective tunneling barrier for the first
excited spin state is indicated by the dashed line.
\label{setup}}
\end{figure}

Our model Hamiltonian matching the  experimental situation in figure~\ref{setup} includes
\begin{equation}
H_0 = \frac{p^2}{2m} + V_\mathrm{B} +  B \mu_\mathrm{B} \tau_z,\label{eqn:modelh}
\end{equation}
which describes  the
tunneling electrons (with a mass $m$ and momentum operator $p$) emanating from the tip and are subject to the magnetic field $B$. They are
coupled to the surface spin subsystem via a spin-dependent term  $V_\mathrm{B}$. The
spin of the tip electrons is described by the vector operators $\hat{\tau} =(\tau_x,\tau_y,\tau_z)$
containing the Pauli matrices in their standard representation. In a similar way,
$\hat{S}=(S_x,S_y,S_z)$ are the operators for the surface spin.
The tunneling barrier $V_\mathrm{B}(x)$ is measured relative to the Fermi levels of
the tip or the sample (for simplicity we assume both to be equal). The last term in equation~\eqref{eqn:modelh} describes
the spin of the tip electrons in the presence of the magnetic field $B$.

\begin{figure}
\includegraphics[width=0.4\textwidth]{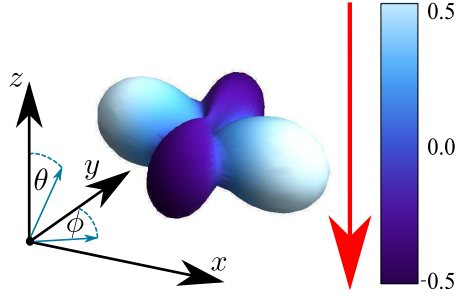}
\caption{The ground state $\chi_1(\theta,\phi)$ in orbital representation. The shape is
given by $|\chi_1(\theta,\phi)|$, whereas the color encodes the real part of the wave
function. The red arrow denotes the expectation value of the spin $\langle \vec{S}
\rangle$.}
\label{groundstate}
\end{figure}

The second subsystem is the surface spin in an anisotropic environment (due to the
broken translational symmetry and other atoms placed nearby). The appropriate
  Hamiltonian for this part  reads~\cite{molecularmagnets1,surfacehamiltonian}
\begin{equation}
\label{hs}
H_{\mathrm{S}} = g \mu_\mathrm{B} B S_z + \delta S^2_z + \varepsilon(S^2_x-S^2_y),
\end{equation}
which indicates that the easy axis is parallel to the (strong) magnetic field.
We have chosen typical values for a rather strong magnetic anisotropy. This does not directly resemble the experimental situation, but serves to illustrate the effects arising from the anisotropic environment more clearly. In particular, we take
$\delta \approx -5$ meV and $\varepsilon \approx 1.5$ meV for the anisotropy constants. The value of the magnetic field is kept with 7 T. The Land\'e
factor $g$ is experimentally~\cite{surfacehamiltonian} found to be approximately 2, which corresponds to the total spin of the free Fe atom ($S=2$). The
Hamiltonian \eqref{hs} has 5 eigenstates $|\chi_\nu \rangle $ with eigenenergies
$E^S_\nu$ (table~\ref{tabeigen} lists their properties). The ground state in orbital representation is depicted in figure \ref{groundstate}.

\begin{table}
\setlength{\doublerulesep}{\arrayrulewidth}
\caption{\label{tabeigen}The spectrum of the adsorbate spin Hamiltonian~\eqref{hs},
referred to the ground state, and the projection of the spin along the magnetic field. The
third column contains the quantum number of  $|m\rangle$ states with the dominant 
contributions to $|\chi_\nu \rangle$; the values are given in the last column.}
\begin{ruledtabular}
\begin{tabular}{cddddd}
\mbox{$\nu$}&\multicolumn{1}{c}{$E^S_\nu-E^S_1$ (meV)}&
\multicolumn{1}{c}{$\langle \chi_\nu | S_z | \chi_\nu   \rangle$}&
m&\multicolumn{1}{c}{$|\langle m | \chi_{\nu} \rangle |^2$}\\
\hline
1 & 0  & -1.507 & -2 & 0.855 \\
2 & 2.006 & 1.498 & 2 &  0.869 \\
3 & 12.121 & -0.088 & -1 & 0.544 \\
4 & 21.156 & 0.088 & 1& 0.544 \\
5 & 22.901 & 0.009 & 0 & 0.944 \\
\end{tabular}
\end{ruledtabular}
\end{table}

The third term on the left side in equation~\eqref{hs} results in a mixing of the different eigenstates $|m\rangle$ of
the free spin in the magnetic field. Thus, the expectation value $\langle S_z \rangle$ in
the ground state deviates from the noninteracting value -2. The ground state
 contains a large amount of  free spin configurations (see table \ref{tabeigen}).
 The first excited state corresponds to the spin pointing in the opposite direction.
Hence, the relaxation $| \chi_2 \rangle \rightarrow |\chi_1 \rangle$  is dominated by the transitions
$|m=+2 \rangle \rightarrow |m=-2 \rangle$.

For the spin-dependent interaction of the tunneling electrons with the adsorbate we adopt
the picture of extended states (tunneling electrons) coupled to a localized magnetic
moment.  In brief, such a process is known to be determined by the Hunds rules and the
local electronic correlation.  E.g. for $s$ states (with energy $\epsilon_{k}$) coupled to
a pinned magnetic impurity (with energy $\epsilon_d$) one may utilize the Anderson
model~\cite{anderson}: $H_{\mathrm{And}} = \sum_{k,\sigma} \epsilon_{k} n_{k \sigma} +
\sum_{\sigma} \epsilon_d n_{d \sigma} + U n_{d \uparrow} n_{d \downarrow} +
H_\mathrm{ds}$, where $\sigma$ labels the spins and $n_{k \sigma}$ ($n_{k \sigma}$) is the
corresponding particle density operator.  $U$ stands for the Coulomb repulsion between
electrons with antiparallel spins. The term $H_\mathrm{ds} = \sum_{k,\sigma} V_{k d}
c^{\dagger}_{k \sigma} c_{d \sigma} +\text{h.c.} $ accounts for the $s\mhyphen d$
hybridization with the mixing matrix element $V_{kd}$. The Anderson Hamiltonian is a model
for a delocalized $s$-band interacting with a single $d$-orbital. As we are interested in
the quantum dynamics of the local magnetic impurity we should map the Anderson model onto
the Kondo Hamiltonian~\cite{andersonkondo} ($H_\mathrm{Kondo} = - J \hat{\sigma}_s \cdot
\hat{\sigma}_d $) with the effective Kondo coupling constant $J$. Tunneling electrons in
the vicinity of the Fermi level are of relevance for the tunneling. This leads to estimate
$J = 2 |V_{k_F d}|^2 \frac{U}{\epsilon_d (\epsilon_d + U)} < 0.$
Typical values~\cite{kondoeffect} are in the order of $|V_{k_F d}|^2 \approx 2$~eV, $U
\approx 8$~eV and $\epsilon_d \approx 5$ eV, meaning that $J \approx 1$~eV.

In STM experiments the tip-sample distance (we take $5$~\AA) is usually not negligible to
the typical extent of $d$-orbitals in Fe adatoms (which is about $1$~\AA). In the
calculations, we use therefore an interaction $H_\mathrm{int} $ of the Kondo type with a
spatial-dependent Kondo coupling
\begin{equation}
H_\mathrm{int} = u_0 w(x) \hat{\tau} \cdot \hat{S}.
\end{equation}
The  $u_0>0$ is the coupling strength and the function $w(x)$ is a
dimensionless form factor defining the width of the interaction region. We choose a
strongly localized function with a width about $1$~\AA.

The total Hamiltonain of the system is thus
\begin{equation}
\label{fullh}
H = H_0 + H_\mathrm{S} + H_\mathrm{int}.
\end{equation}
The total effective potential experienced by the tunneling electrons $V_\mathrm{B}(x) +
\langle \chi_{\nu}, \uparrow|H_\mathrm{int} |\chi_{\nu}, \uparrow \rangle$ is thus
different for the ground state $\nu=1$ and the first excited state $\nu=2$, as sketched in
figure~\ref{setup}~(b) (the tip spin is $\uparrow$ in both cases). The anti-ferromagnetic
coupling of the tip spin and the surface spin effectively raises the tunneling barrier for
the tip electrons, for they have to overcome the additional spin-interaction energy. On
the other hand, any excitation of the surface spin lowers the effective potential barrier
(c.~f. solid and dashed lines in figure~\ref{setup}~(b)). Therefore the excitation
$|\chi_1 \rangle \rightarrow |\chi_2 \rangle$ is favored as long as the applied voltage is
higher than the first excitation energy, i.~e. if $eV > E^S_2-E^S_1$. For a smaller
voltage this excitation is not reached energetically.  All higher excited states also
lower the effective potential barrier, but the effect is maximal for $|\chi_2
\rangle$. This is reflected in the enhanced excitation probability for the first excited
state.

If we on the other hand consider the case of $|\!\downarrow\rangle$-electrons beeing
emitted from the tip, the effective potential $V_\mathrm{B}(x) + \langle \chi_{\nu},
\downarrow|H_\mathrm{int} |\chi_{\nu}, \downarrow \rangle$ undergoes a change of the sign
in the last term. Therefore, $\langle \chi_{1},\downarrow|H_\mathrm{int} |\chi_{1},
\downarrow \rangle$ becomes a negative contribution and thus enhances the tunneling
current, whereas $\langle \chi_{2},\downarrow|H_\mathrm{int} |\chi_{2}, \downarrow
\rangle$ acts as an additional barrier. The other consequence is the smaller excitation
probability because the transition $|\chi_1\rangle \rightarrow |\chi_2 \rangle$ does not
reduce the tunneling barrier. Physically, this picture reflects the anti-ferromagnetic
coupling again.

We remark that the existence of a direct transition $|\chi_1 \rangle \rightarrow |\chi_2
\rangle$ is governed by the transition matrix element $ \langle \chi_{\mu},
\uparrow|H_\mathrm{int} |\chi_{\nu}, \uparrow \rangle$ in the surface spin space, which is
connected to the anisotropy constant $\varepsilon$. For the rather large value of
$\varepsilon \approx 5$~meV, the direct excitation $|\chi_1 \rangle \rightarrow |\chi_2
\rangle$ is the dominant process~\cite{epsdiscussion}.

\section{Scattering formulation \label{secscattering}}

In a wave-packet scattering picture of the process depicted in figure~\ref{setup} one
identifies an undisturbed part $H_\mathrm{free} = p^2/(2m) + g_e \mu_\mathrm{B} B \tau_z +
H_\mathrm{S}$ and a scattering part $H_\mathrm{sc} = V_\mathrm{B}(x) + H_\mathrm{int}$.
The unperturbed (asymptotic) scattering states are cast as ($\pm$ stand for
incoming/outgoing waves) $|\Phi^{\pm}_{\nu,\tau} \rangle = | \pm k \rangle \otimes
|\chi_\nu \rangle \otimes |\tau \rangle$ where $|\chi_\nu \rangle$ ($|\tau\rangle$) is the
spin ground state of the surface (tip).  $| \pm k \rangle$ is the orbital part of the wave
with a wave vector $k$ which is determined by the direction and by the amplitude of the
applied bias voltage $V$, i.e.  $k= \sqrt{2 m eV / \hbar^2}$ (and $e$ is the electron
charge). By propagating $|\Phi^+_{\nu,\tau} \rangle$ the state of the eigenstates
$|\psi^\pm_{\nu,\tau} \rangle$ of the full Hamiltonian can be obtained (up to a phase)
~\cite{scatteringtheory1,scatteringtheory2,gellmannlow}, i.e.  $|\psi^+_{\nu,\tau} \rangle
= U(0,-\infty) |\Phi^+_{\nu,\tau} \rangle$ and $|\psi^-_{\nu,\tau} \rangle = U(+\infty,0)
|\Phi^-_{\nu,\tau} \rangle$, where $U(t,t_0)$ is the propagator in the interaction
representation.  Tunneling is governed by the probability amplitude to scatter from the
incoming eigenstate $|\psi^+_{\mu,\sigma} \rangle$ into the outgoing state
$|\psi^-_{\nu,\tau}\rangle$ is given by the scalar product $\langle \psi^-_{\nu,\tau} |
\psi^+_{\mu,\sigma} \rangle = \langle \Phi^-_{\nu,\tau} | U(+\infty,-\infty )|
\Phi^+_{\mu,\sigma} \rangle$. The last equation defines the on-shell S-matrix $S =
U(+\infty,-\infty)$. Evaluating the S-matrix element
\begin{equation}
\label{defsmatrix}
S^{\sigma \tau}_{\mu \nu} =\langle \Phi^-_{\nu,\tau} | S| \Phi^+_{\mu,\sigma} \rangle
\end{equation}
we find the tunneling probability to scatter from the tip electron with a spin state
$|\tau\rangle$ into $|\sigma\rangle$ and from the surface spin state $|\chi_\nu \rangle$
into $|\chi_\mu \rangle$ by taking the absolute value of the S-matrix (equation~\eqref{defsmatrix}).
For a given initial spin-spin density matrix $\rho_0$ (of the dimension 10):
\[
\rho_0 = \rho^{tip}_0 \otimes \rho^S_0,
\]
we obtain the total tunneling probability $T(V)$ and, thus, the normalized density matrix
elements by
\begin{equation}
\rho^{\sigma \tau}_{\mu \nu} = \frac{1}{T(V)}
\langle \Phi^-_{\nu,\tau} |U(+\infty,0) \rho_0 U(0,-\infty)| \Phi^+_{\mu,\sigma} \rangle.
\end{equation}
The probability to scatter into a specific channel $(\nu,\sigma)$ is given as
\begin{equation}
\label{channelprobability}
T_{\nu \tau}(V) =
\langle \Phi^-_{\nu,\tau} |U(+\infty,0) \rho_0 U(0,-\infty)| \Phi^+_{\nu,\tau} \rangle,
\end{equation}
Whereas the total tunneling probability is   $T(V) = \sum_{\nu,\tau} T_{\nu
\tau} (V) = \text{Tr} \rho$.

In the setup figure~\ref{setup} we are interested in the surface spin dynamics only.
 Hence, we trace out the spin $|\sigma\rangle$ of the tunneling tip electron. This yields
 the reduced density matrix of the surface spin. Furthermore, we fix the initial spin
 state of the tip electron as $|\tau\rangle = |\!\uparrow\rangle$ as determined by the
 polarization of the tip and $|\chi_\nu \rangle = |\chi_1 \rangle$, i.~e. the surface spin
 is in the ground state when we switch on the voltage. We simplify the notations and write
 the resulting density matrix as $\rho^S_{\mu \nu}$(V). The population of the final
 surface spin state $|\chi_\nu \rangle$ as a function of the applied voltage is then given
 by $P^S_{\nu}(V) = \rho^S_{\nu \nu}(V)$, whereas the phase information can be extracted
 from the off-diagonal elements.

The assumption $|\tau\rangle = |\uparrow \rangle$ corresponds to the case of a full  polarization of the tip.
 A partial polarization $\beta$ can be included in the formalism via the initial
spin density matrix of the tip, i.e. 
\begin{equation}
\rho^{tip}_0 = \frac12 \begin{pmatrix} 1+\beta &0 \\ 0 & 1-\beta \end{pmatrix}
\end{equation}
 For clarity however,  we employ   
 $\beta=1$ throughout this work and comment on the change of the results for 
 $\beta <1$.

To calculate the individual S-matrix elements we utilize the wave packet
approach~\cite{wavepacket1,wavepacket2}. This method exploits the fact that in the
Schr\"odinger picture the full eigenstates can be expressed as a superposition of
propagated wave packets, i.e.
\begin{equation}
\label{fullpsi}
| \psi^{\pm}_{\nu,\tau} \rangle = \frac{1}{2 \pi \hbar \eta^\pm_{\nu \tau}}
\int^{\infty}_{-\infty} dt \ e^{-i H t / \hbar} e^{i E t / \hbar} |\phi^{\pm}_{\nu,\tau} \rangle,
\end{equation}
where the normalization factor $\eta^\pm_{\nu \tau}$ is defined by
\begin{equation}
\label{norm}
\eta^\pm_{\nu \tau} = \sqrt{\frac{m}{2 \pi \hbar k}}
\int^{\infty}_{-\infty} dx \ g_\pm(x) e^{i k x}.
\end{equation}
The wave vector $k$ for $\eta^+_{\nu \tau}$, i.~e. for the incoming waves, is given by $k=
\sqrt{2 m eV / \hbar^2}$, whereas for the outgoing state, the wave vector may be reduced
due to inelastic tunneling. In this case, we set $k = \sqrt{2 m (e V + E^S_1 +
E^{\mathrm{tip}}_\tau - E^S_{\mu} - E^{\mathrm{tip}}_\sigma) / \hbar^2}$ if the term under
the square root is positive. Here $E^{\mathrm{tip}}_\gamma$ ($E^S_{\gamma}$) is the energy of the
tip (surface) with a spin state $\gamma$.

The wave packets $|\phi^{\pm}_{\nu,\tau}\rangle$ are superpositions of full
eigenstates. However, by placing the wave packets far away from the interaction region
$U(t,t_0)$ becomes the unit operator in which case the packets
by composed of the eigenstates of $H_\mathrm{free}$, i.~e. we can write $|\phi^\pm_{\nu \tau} \rangle $
as the product state $|g_{\pm}\rangle |\chi_{\nu}\rangle |\tau \rangle$ with a Gaussian
function $g_{\pm}$. The subscripting $\pm$ now corresponds to centering the Gaussian to the
left and to the right of  the interaction region.

Using the spectral representation (equation~\ref{fullpsi}) we find the S-matrix elements as
\begin{equation}
\label{smatrix}
S^{\sigma \tau}_{\mu \nu} = \frac{1}{2 \pi \hbar (\eta^-_\mu)^* \eta^+_\nu}
\int^{\infty}_{-\infty} dt
\langle \phi^-_{\mu,\sigma} | e^{-i H t / \hbar} | \phi^+_{\nu,\tau} \rangle.
\end{equation}
The formula~\eqref{smatrix} is then evaluated by (i) propagating  the initial product
state, (ii) calculating the overlap with the final state on the right side of the barrier,
and (iii) by performing a Fourier transformation.

\section{Spin state population \label{secspinstatepop}}

\begin{figure}
\includegraphics[height=0.6\textheight]{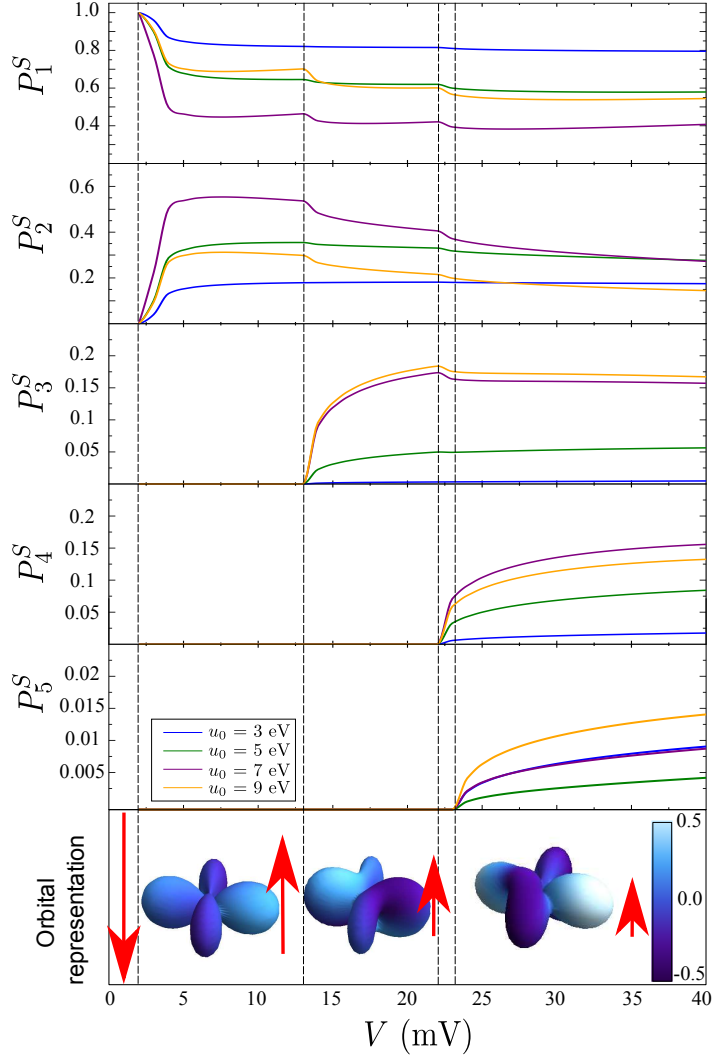}
\caption{The population $P^S_{\nu}$ for all spin states of the surface-deposited structure
as a function of the applied voltage for various coupling strengths $u_0$. The black
dashed lines indicate when the voltage is high enough as to excite the surface-deposited
spin states obeying the energy condition $E^S_\nu + E^{\mathrm{tip}}_{\sigma}
-E^{\mathrm{tip}}_{\uparrow}- E^S_1-eV>0$. The bottom row depicts the voltage-dependent
wave functions in orbital representation (for $u_0=7$~eV), evaluated for the highest
voltage between two black dashed lines. The color coding is as in
figure~\ref{groundstate}. The red arrow indicates the expectation value of the spin. Note
that when exceeding the first threshold voltage, the spin direction flips, as it is
primarily connected to the transition $| m= +2 \rangle \rightarrow |m = -2
\rangle$. \label{population1}}
\end{figure}

We now proceed with computing the population $P^S_{\nu}$ of the surface spin eigenstates
$|\chi_\nu\rangle$ as a function of the applied voltage $V$ for different coupling
constants $u_0$. We assume the surface spin to be initially in the ground state and the
spin of the tip's electron is being aligned with the magnetic
field. Figure~\ref{population1} depicts the dependence of the population of all surface
spin states. The energy of the tunneling electrons ($e V$) determines which inelastic
scattering processes are possible. For voltages below the excitation threshold the
corresponding spin state can not be excited due to the energy conservation. For the other
transitions, one has to take the spin of the tip's electron into account, as well. By
evaluating the spin-coupling $\hat{\tau} \cdot \hat{S}$ in the basis of product states one
finds that the transitions $|\chi_1 \rangle \rightarrow |\chi_3 \rangle$ and $|\chi_1
\rangle \rightarrow |\chi_4 \rangle$ are only allowed if accompanied by spin flip
processes. Other transitions require the conservation of the spin direction. We can, thus,
define specific voltages $V_\nu$, where new tunneling channels open, i.~e. $e V_\nu =
E^S_\nu + E^{\mathrm{tip}}_{\sigma} -E^{\mathrm{tip}}_{\uparrow}- E^S_1$. For $\nu \neq
3,4$ we obtain $e V_\nu = E^S_\nu - E^S_1$ whereas for $\nu = 3,4$ $eV_\nu = E^S_\nu +
E^{\mathrm{tip}}_{\downarrow} -E^{\mathrm{tip}}_{\uparrow}- E^S_1$. The voltages $V_\nu$
are shown in figure~\ref{population1} by vertical dashed lines.

The basic trend of the voltage-dependence of the population is to reduce the coupling
energy by exciting the surface spin. As long as the energy of the tunneling electrons is
high enough, the population of the ground state decreases. For voltages in the interval
between the first two dashed lines, the first excited state is the only accessible
tunneling channel, so  that the population is solely transferred to $|\chi_2
\rangle$. The efficiency of this transition is determined by the coupling constant
$u_0$. If the voltage increases further, the next tunneling channels open and  the
spin interaction energy can again be lowered by exciting the surface spin. Therefore,
$P^S_{2}$ decreases and $P^S_{3}$ rises by the same amount. The situation is
similar between the last two dashed lines, where $P^S_{3}$ is decreased in order to
increase $P^S_{4}$. The highest state $|\chi_5 \rangle$ corresponds to a very small
projection parallel to the magnetic field as the major contribution is $|m=0 \rangle$ (see
table~\ref{tabeigen}). The transition probability is, thus, much smaller than a transition
into the other states. For the surface spin it is therefore more convenient to maintain a
higher population of $|\chi_4 \rangle$ instead of decreasing $P^S_{4}$ when the
voltage exceeds the last threshold.

\begin{figure}
\includegraphics[width=0.7\columnwidth]{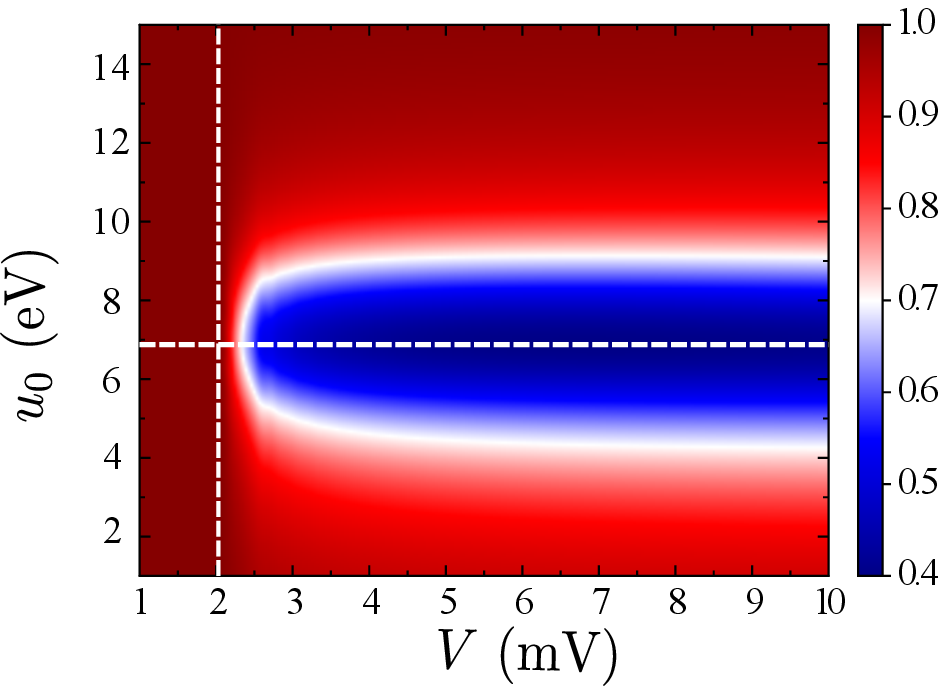}
\caption{The population $P^S_{1}$ in the ground state as a function of the voltage $V$
and the coupling strength $u_0$. The voltage is bounded such that the only accessible
tunneling channel is the first excited spin state of the adatom. the vertical white
dashed line indicates the energy difference $E^S_2-E^S_1$. We find that the depopulation
process of the ground state is particularly efficient for $u_0 \approx 7$ eV. For stronger
or weaker coupling strengths, the excitation probability decreases. This can be understood
by a simple analytical model (see appendix). The prediction of this consideration for the optimal value
$u^c_0$ is shown by the horizontal white dashed line. \label{densplot}}
\end{figure}

For larger coupling constants $u_0$ the probability to scatter to the excited states of
the surface spin increases. However, this holds true as long as $u_0 < u^c_0$, where
$u^c_0$ is about $7$ eV. For larger $u_0$, the transition efficiency decreases again. In
order to reveal the dependence on $u_0$, we calculated the population of the ground state
as a function of $V$ and $u_0$ (figure~\ref{densplot}).  Since the effects are especially
pronounced for the transition $|\chi_1 \rangle \rightarrow |\chi_2 \rangle$, we will only
consider the corresponding voltage interval.  As it turns out, the excitation of the
surface spin is only probable for certain values of the coupling constant. For large
$u_0$, the excitation probability even approaches zero.

In order to explain this behavior, we have to apply a further simplification that renders
possible an analytical treatment (see appendix). Analyzing the simplified model we infer
that
\begin{equation}
\label{u0optimal}
u^{c}_0 = \frac{\hbar^2 q} {2 m x_0 C_{22}}
\end{equation}
with $C_{22} = \langle \chi_2 | S_z |\chi_2 \rangle$ and $q = \sqrt{2 m V_B /\hbar^2}$ for
the optimal value of the coupling strength. Physically, the occurrence of this optimal
value can simply be interpreted as a resonance phenomenon of scattering on the narrow
quantum well of the spin-spin-interaction potential. The value of $u^c_0$ is represented
by the horizontal white dashed line in figure~\ref{densplot}.

Generally, for values for the tip polarization $\beta <1$, the excitation probability
decreases. We have already noted that the transitions induced by the tunneling electrons
with the opposite spin direction ($|\tau \rangle = |\!\downarrow \rangle$) are weaker in
comparison to $|\! \uparrow\rangle$-electrons. For $\beta$ decreasing from $\beta=1$, the
strong ($|\!\uparrow \rangle$) and the weak ($|\!\downarrow\rangle$) contribution are
mixed together~\cite{finitebeta}. Despite from beeing reduced to a smaller intervall of
the probability, the shape of the curves in figure \ref{population1} remains the same for
realistic values of $\beta \approx 30$ \%.

The resonance phenomenon arising from the quantum well of the effective potential becomes
also less visible for a smaller tip polarization, as the quantum well is inverted to a
barrier for electron spins of the opposite direction. The value of $u^c_0$ is however
hardly influenced.

\section{Relaxation dynamics \label{secpumpprobe}}
STM experiments are not only suitable for determining the transport properties, but
also offers an insight in the relaxation dynamics of the excited surface spin by using a
pump-probe scheme. The time resolution is not sufficient to resolve the precession or
the transitions between neighboring $|m\rangle$ states, but it is  still high enough to measure the
relaxation from $|m=+2 \rangle$ to $|m=-2 \rangle$. This corresponds in a good
approximation to the transition $|\chi_2 \rangle \rightarrow |\chi_1\rangle$.

The idea of determining the relaxation dynamics is to apply a pump voltage pulse in order
to excite the surface spin, i.~e. to increase the population $P^S_{2}$ and to detect the
time evolution by a second probe pulse. The voltage of the latter has to be chosen
according to the excitation threshold. We set the maximal values $V^{(0)}_{\mathrm{pump}}
= 3$ mV and $V^{(0)}_{\mathrm{probe}} = 1$ mV.  This specific value for the pump voltage
is sufficient to induce an excitation of the adsorbate spin. If the excitation has to take
place with the help of intermediate states, e.~g. if the transition matrix element does
not allow for the direct pathway $|\chi_1 \rangle \rightarrow |\chi_2 \rangle$, the pump
voltage has to be chosen higher according to the energy of these intermediate states.  We
note that in our case the sum of $V^{(0)}_{\mathrm{pump}}$ and $V^{(0)}_{\mathrm{probe}}$
is much smaller than the voltage $V_3$ needed to excited the third or higher excited
states. The tip spin can not flip for the transition $|\chi_1 \rangle \rightarrow |\chi_2
\rangle$, hence we can reduce our considerations to a two-level system (TLS).

\begin{figure}
\includegraphics[width=0.6\columnwidth]{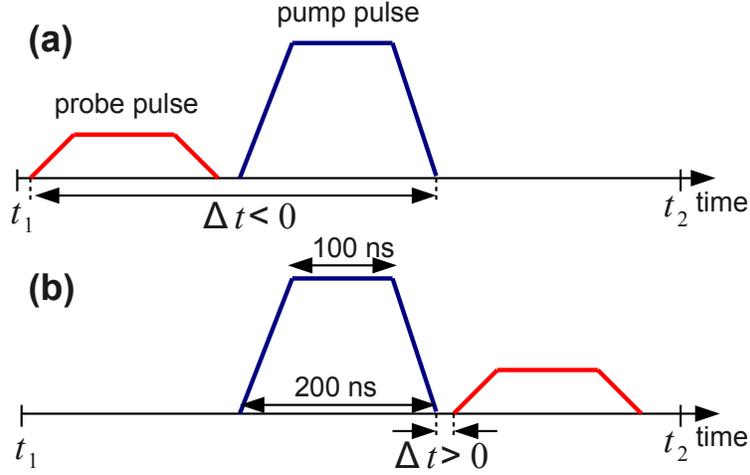}
\caption{A sketch of the pump-probe arrangement of the voltage pulses. In the case (a) the probe
pulse advances the pump pulse characterized by $\Delta t <0$ and the cross correlation of
both pulses is zero. On the other hand, by varying $\Delta t>0$ as in (b) one can
determine the relaxation time for the transition $|\chi_2 \rangle \rightarrow
|\chi_1\rangle$. The probe and the pump pulse both have a duration of $200$ ns with the
linear rise/fall intervals of $50$ ns.
\label{pulses}}
\end{figure}

The experimentally accessible quantity is the number of tunneling electron $N$ during the
time interval containing the pump and the probe pulses shifted against each other by a
time delay $\Delta t$ as displayed in figure~\ref{pulses}.

For large $|\Delta t|$ and $\Delta t <0$ (\ref{pulses}a), there is no overlap of
the pump and probe pulse. The number of tunneling electrons is therefore just the sum of
the electrons transmitted during probe pulse $N_{\mathrm{probe}}$ and pump pulse
$N_{\mathrm{pump}}$ separately. This number of electrons will be denotes as $N_0 =
N_{\mathrm{pump}}+ N_{\mathrm{probe}}$. For smaller $|\Delta t|$ but still negative
$\Delta t$, the pump and the probe pulse have a finite overlap, i.~e. there is a certain
time interval where the effective voltage is $V_{\mathrm{pump}} + V_{\mathrm{probe}}$. If
the current-voltage characteristic is nonlinear, the $N$ will be different from $N_0$
because of the correlation of probe and pump pulse. In our case, the nonlinearity mainly
arises from the coupling Hamiltonian $H_\mathrm{int}$.

For  $\Delta t > 0$ the pump and the probe pulses have again no overlap, but the pump pulse
depletes the population of the ground state. The following probe pulse can only
access the ground state scattering channel, hence the number of transmitted electrons is
lower than $N_0$. With $\Delta t$ increasing, there is more time for the surface spin to
relax back to the ground state, so $N$ approaches $N_0$ again for a large $\Delta t$.

\begin{figure}
\includegraphics[width=0.6\columnwidth]{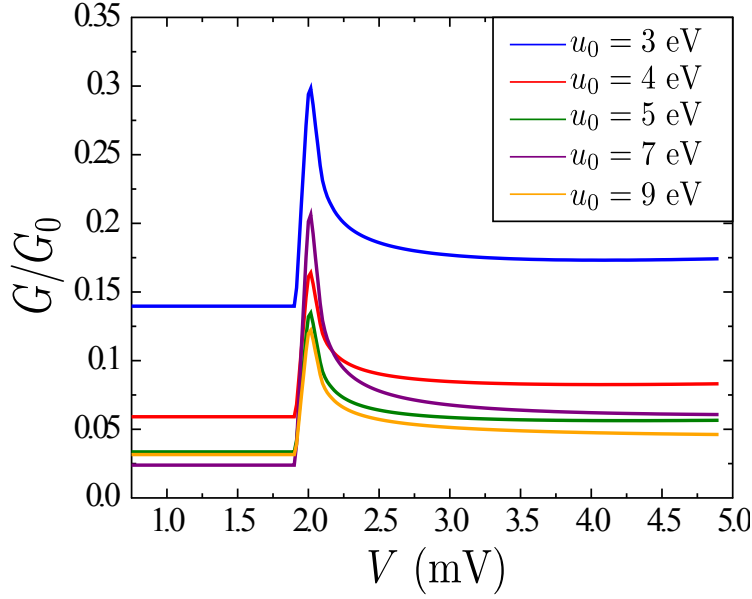}
\caption{The conductance of the magnetic tunneling junction for various coupling
strengths $u_0$. The value of $G$ has been normalized to the situation where $u_0=0$,
in which case the conductance is almost constant.\label{conductance}}
\end{figure}

For a quantitative analysis, we note that all quantities are only parametrically
time-depended, since the time-dependence of $V_{\mathrm{pump}}(t)$ and
$V_{\mathrm{probe}}(t)$ is very slow on the typical time scales of the system. This means
we can use the results of the steady-state considerations from the
section~\ref{secscattering} and \ref{secspinstatepop}. From the tunneling probability for
the different scattering channels $T_{\nu \sigma}(V)$ we compute the corresponding
currents as $j_{1,2} = (e\hbar k_{1,2}/m )T_{(1,2)\uparrow}(V)$ with $k_1 = \sqrt{2 m e V
/ \hbar^2}$ and $k_2 = \sqrt{2 m (eV + E^S_1-E^S_2)/\hbar^2}$ for $eV > E^S_2-E^S_1$ and
$k_2 = 0$ otherwise. The total current is expressed as the sum $j = j_1 + j_2$
proportional to the total number of tunneling electrons $dN$ in the given time interval
$dt$. As the calculations are based on the quasi-static regime, it is still possible to
define the conductance $G(V) \propto d j / dV$, which then is parametrically
time-dependent, as well. As we have pointed out in section~\ref{secspinstatepop}, the
excitation of the state $|\chi_2 \rangle$ is favored as long as the voltage is
sufficiently high. For this reason, the average slope of $j_2(V)$ is larger than the
average slope of $j_1(V)$, i.~e. the conductance jumps to a higher value if $eV$ exceeds
$E^S_2-E^S_1$. The nonlinear conductance profile is depicted in figure~\ref{conductance}.

We note the population of the ground state and the first excited state evolve only
adiabatically with resultant voltage $V_{\mathrm{pump}}(t) + V_{\mathrm{probe}}(t)$,
because the excitation $|\chi_1\rangle \rightarrow |\chi_2\rangle $ occurs instantly on
the time scale of the voltage pulses. With the surface spin driven into the excited state,
it undergoes i) a precession (the characteristic time is about 2~ps) and ii)
relaxes. Since the computation of $N$ involves an integration over the time interval of
several hundred nanoseconds, the oscillatory part due to the precession does not play a
role. We can, therefore, describe the dynamics of the surface spin in the case of a
falling voltage by using the Bloch equations including damping, but at a zero
frequency. The time scale for this relaxation process, chosen according to the
experiment~\cite{experiment}, is again very slow for our system.

For a pump-probe delay $\Delta t <0$ such that the probe pulse still advances the pump
pulse, we therefore obtain
\begin{equation}
N = \frac{1}{e L} \int dt~j(V_{\mathrm{pump}}(t) + V_{\mathrm{probe}}(t))
\end{equation}
with $j(V) = e\hbar (k_1 T_{1 \uparrow}(V)+k_2 T_{2 \uparrow}(V))/m$. For $\Delta t > 0$
on the other hand, there is no overlap of the two pulses and hence the number of electrons
transmitted due to the pump pulse is given by $N_{\mathrm{pump}} = \frac{1}{e L} \int dt ~
j(V_{\mathrm{pump}}(t))$. The probe pulse can only access the ground state $|\chi_1
\rangle$. The tunneling probability and equivalently the current are in this case
proportional to the population $P^S_{1}$, which we obtain by calculating
$P^S_{1}(V^{(0)}_{\mathrm{pump}} + V^{(0)}_{\mathrm{probe}}) < 1$ and then computing the
increase of $P^S_{1}(t)$ back to 1 due to the relaxation. The number of tunneling
electrons due to the probe pulse $N_{\mathrm{probe}}=\frac{1}{e L} \int dt~P^S_{1}(t)
j_1(V_{\mathrm{probe}}(t))$ is, thus, lowered in comparison to the probe pulse advancing
the pump pulse.

\begin{figure}
\includegraphics[width=0.6\columnwidth]{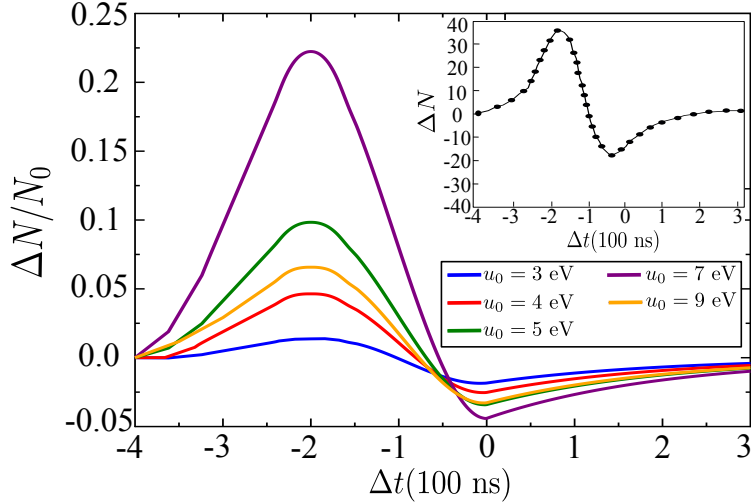}
\caption{The relative number of the transmitted electrons with respect to $N_0$ as a
function of the delay $\Delta t$ of pump and probe pulses for various interaction
strengths $u_0$. The curve has a peak at $\Delta t = - 200$ ns which refers to a complete
overlap of the probe and the pump pulses. Since the population in the ground state is
lowered by the excitation of the surface spin, $N-N_0$ becomes negative for $\Delta t>0$
and approaches zero again when the excited state relaxes back into the ground state.  Our
calculations, thus, confirm that threshold phenomena may also arise under different
excitations: $|\chi_2\rangle$ is excited directly. For comparison, the inset shows the
experimental result from Ref.~\onlinecite{experiment} where the same final state is
reached by a sequence of excitations with $\Delta m=1$.\label{ncurve}}
\end{figure}

Figure~\ref{ncurve} shows the total dependence of $N$ on the time delay $\Delta t$. Again
the curves in figure~\ref{ncurve} evidence a strong dependence on the coupling of the
tunneling electron to the local moment $u_0$. The peak at $\Delta t = -200$ ns
(corresponding to a complete overlap of the pump and the probe pulses) is due to the
nonlinearity of current-voltage characteristic, as
$j(V^{(0)}_{\mathrm{pump}}+V^{(0)}_{\mathrm{probe}}) > j(V^{(0)}_{\mathrm{pump}}) +
j(V^{(0)}_{\mathrm{probe}})$. The magnitude of this effect can be explained by means of
the the conductance profile (figure~\ref{conductance}). For small values of $u_0$, the
relative jump of the conductance is small, so $j(V)$ is nearly linear,
i.~e. $j(V_{\mathrm{pump}}(t)+V_{\mathrm{probe}}(t)) \gtrsim j(V^{(0)}_{\mathrm{pump}}) +
j(V^{(0)}_{\mathrm{probe}})$. Hence, $N$ only slightly exceeds $N_0$ when both pulses
coincide. For this reason, the first peak in figure~\ref{ncurve} is small for smaller
$u_0$. The maximum becomes more pronounced when approaching the optimal value of $u_0$ for
the depopulation of the ground state. From the conductance profile
(figure~\ref{conductance}) we conclude that the relative jump of $G$ when exceeding the
threshold voltage attains the highest value for $u_0 = u^c_0$. The nonlinearity then plays
a dominant role, resulting in an especially strong increase of $N$ with respect to
$N_0$. For larger coupling strength, the current-voltage dependence resembles again the
linear case. Note that the averaged conductance is much smaller for $u_0 > u^c_0$ than in
the case $u_0 < u^c_0$. The minimum at $\Delta t = 0$ is caused by the reduction of the
ground state population $P^S_{1}$ due to the advancing pump pulse. We have already
discussed the efficiency of this process (figure~\ref{densplot}). For values $u_0 \ll
u^c_0$ or $u_0 \gg u^c_0$, $P^S_{1} \lesssim 1$ holds. Hence, $N_{\mathrm{probe}}$ and for
this matter $N$ decrease only by a small amount. The depth of the minimum is maximal
for~$u_0 = u^c_0$.

We remark that the relative jump in the conductance profile also decreases for $\beta
<1$. The tunneling channel connected to the excitation $|\chi_1 \rangle \rightarrow
|\chi_2 \rangle$ is much less effective for $|\! \downarrow \rangle$-electrons emitted
from the tip. Hence, the major contribution for this electrons arises from the adsorbate
spin remaining in the ground state. Hence, the conductance for voltages $eV > E^S_2 -
E^S_1$ differs only slightly from the value left to this threshold. This effect again
shows a linear dependence on the tip polarization. The total value of the conductance on
the other hand increases with smaller $\beta$, which is due to the reduction of the
effective potential for $|\! \downarrow \rangle$-electrons with the adsorbate spin in the
ground state.

Similarly to the discussion in section \ref{secspinstatepop}, the special features with
respect to $u^c_0$ become less visible if we reduce $\beta$. Hence, the height of maximum
in figure \ref{ncurve} decreases. This can be explained in terms of the conduction profile
which more and more resembles the linear case. On the other hand, the transition of the
surface spin to the first excited state effectively blocks the $|\! \downarrow
\rangle$-electrons from the tips and therefore reduces the current more than in the case
$\beta=1$. We thus conclude that the minimum becomes deeper for a realistic tip
polarization.

\section{Conclusions}
We analyzed theoretically the  dynamics of a surface-deposited localized magnetic moment,
triggered by a spin-polarized current pulse from an STM tip in the presence of a magnetic
field. We showed and explained how and why the spin population of the surface-deposited
structure can be manipulated by applying appropriate voltages and investigated the
dependence on the coupling strength $u_0$ of the tunneling electrons to the localized
moments.  A simplified analytical model allowed us to predict a value of $u_0$ for which
the excitation probability of the surface spin is particularly high. The obtained value is
in excellent agreement with the numerical calculations.

The relaxation was incorporated by using the Bloch equations, as our model does not
provide a relaxation mechanism. For future work, this point deserves a special
consideration. In this respect, the main advantage of the Cu$_2$N-overcoat is the magnetic
decoupling of the adatom and the substrate, the excitation of magnons is strongly
suppressed leading to prolonged relaxation times. In general, we suggest the spin-orbit
interaction, that transfers the energy to the substrate, as the major dissipation channel.

The above treatment is based on an adiabatic approximation in the sense, that the voltage
pulses durations are very long on the time scale of the electron and spin dynamics, which
is a very well justified assumption for the experiments under consideration
(figure~\ref{setup}).  In future experiments one may envisage however to go beyond the
relaxation dynamics with the aim to image the precession of the adsorbate spin as well.
In this case picosecond pulses are required.  Such pulses have already been utilized for
tracing the ultrafast spin dynamics in a proof of principle
experiments~\cite{slac1,slac2}. Alternatively, one may consider laser-pulse induced
tunneling currents, as in the recent experiments~\cite{atto,atto2}.  Generally, the
characteristic times for the tunneling process (some femtoseconds) and the spin dynamics
(picoseconds) are still very different. Hence, the dynamics driven by the voltage pulses
with picosecond durations can be adiabatically decomposed into (i) the time evolution of
the electron states, (ii) the electron pulse driving the surface spin, and (iii) an
adiabatic dependence of the current on the surface-deposited spin dynamics (which images
of the spin dynamics).  This time scale separation is the key for future experiments and
calculations on road to full insight into the spatiotemporal evolution of the quantum spin
dynamics.

\section{Appendix: Analytical model\label{anamodel}}

In section~\ref{secspinstatepop} that for the parameters of interest the excitation
probability exhibits a maximum for $u_0 \approx 7$ eV. To explain the specific value of
the maximum and to gain further insight in the underlying physics, we introduce here some
reasonable simplifications that allow an analytically solvable model.  Since the
scattering channels are limited to the first two states of the surface-deposited magnetic
moment and the tip electron spin can not flip within this transition, we can reduce our
considerations to a two-channel model. The operator $\hat{\tau} \cdot \hat{S}$ can be
replaced by $S_z$ as the electron spin is fixed in the $z$-direction. For small voltages,
the gradient of the potential barrier can be neglected. We can, thus, assume a rectangular
shape of the barrier. The function $w(x)$ describing the spin-interaction area is
localized around the Fe atom. Hence we approximate the coupling Hamiltonian by
\begin{equation}
\label{approxhint}
H_\mathrm{int} = u_0 C ~ \delta\left(\frac{x-L}{x_0}\right),
\end{equation}
where $x_0$ is a typical length associated with $w(x)$, $L$ is the width of the vacuum
barrier and $C$ is the remaining coupling matrix that contains the  transition amplitudes,
i.~e. $C_{\alpha \beta} = \langle \chi_{\alpha} | S_z | \chi_{\beta} \rangle$.

The wave functions $\psi^+_\nu(x)$ are then linear combinations of plane waves $e^{\pm i k
(x-L)}$ for $x <0$ or $x>L$ and $e^{\pm q (x-L)}$ for $x \in (0,L)$. The respective wave
vector follows as $k_\nu = \sqrt{2 m (e V +E^S_1-E^S_\nu) /\hbar^2}$, whereas
$q_\nu=\sqrt{2 m (V_B-e V -E^S_1+E^S_\nu) /\hbar^2}$. Since the height of the vacuum
barrier $V_B$ is much greater than the electron energy, we can approximately set $q_\nu =
q = \sqrt{2 m V_B /\hbar^2}$. For $0<x<L$ we write $\psi^+_\nu(x) = \alpha^{<}_{\nu} e^{q
x} + \beta^{<}_{\nu} e^{-q x}$ and for $x>L$ we define $\psi^+_\nu(x) = \alpha^{>}_{\nu}
e^{i k_\nu x} + \beta^{>}_{\nu} e^{-i k_\nu x}$. The coefficients are linked by the
corresponding transfer matrices. Imposing the standard matching conditions for the
wave-function and its derivative we obtain for the reflection coefficient of the first
excited spin state in the interval $(0,L)$:
\begin{equation}
\beta^<_2 = -\alpha^>_1 \frac{m}{\hbar^2} \frac{ x_0 u_0 C_{12}}{q} +
\alpha^>_2  \left( \frac{i k_2}{2 q}+
\frac{\hbar^2 q -2 m x_0 u_0 C_{22} }{2 \hbar^2 q}\right),\label{transeq}
\end{equation}
where $>$ and $<$ labels the coefficients to the left and to the right of $x=L$. We can
furthermore neglect the term $k_2/q$ because of the small voltages. The remaining terms
are then real, so that the absolute value of $\beta^<_2$ is minimal if the right side in
equation~\eqref{transeq} is minimal. Hence, $\alpha^>_1$ becomes small near the optimal
value $u^c_0$. We conclude that $\beta^<_2$ is small for $q = 2 m x_0 u_0 C_{22} /
\hbar^2$, yielding for the optimal value $u^c_0 = \hbar^2 q / (2 m x_0 C_{22})$.

\begin{acknowledgements}
Discussions and consultations with Markus Etzkorn are gratefully acknowledged.
J.B. and Y.P. thank the DFG for financial support through SFB 762.
\end{acknowledgements}

\providecommand{\newblock}{}


\begin{thebibliography}{10}
\bibitem{epsdiscussion} By evaluating the transition element $\langle \chi_\mu, \tau|
\hat{\tau} \cdot \hat{S} | \chi_\nu, \sigma \rangle$, we found that the direct excitation
to the first excited state is always possible for $\varepsilon >0$. Furthermore, the
direct pathway turns out to be very efficient for the parameters used in this work. This
changes if the value of $\varepsilon$ decreases, i.~e. $\langle \chi_1, \uparrow|
\hat{\tau} \cdot \hat{S} | \chi_2, \uparrow \rangle$ becomes very small, so that stepwise
excitations over intermediate states start to dominate.
\bibitem{finitebeta} One can readily show that the population $P^S_\nu(V)$ of the
eigenstates of $H_\mathrm{S}$ change according to $P^S_\nu(V,\beta) =
\frac{1+\beta}2P^S_\nu(V,\tau=\uparrow) + \frac{1-\beta}2P^S_\nu(V,\tau=\downarrow)$ when
taking a finite tip spin polarization $\beta$ into account. This formula connects the
probabilities with the initial condition for $|\tau\rangle$ in a linear way.
\end{thebibliography}
\end{document}